\documentclass{mn2e}
\usepackage{amssymb,amsmath}
\usepackage[numbers,sort&compress]{natbib}

\bibpunct{[}{]}{,}{n}{}{;}

\newcommand{\be}{\begin{equation}}
\newcommand{\ee}{\end{equation}}
\newcommand{\ben}{\begin{eqnarray}}
\newcommand{\een}{\end{eqnarray}}

\newcommand{\lb}{\left(}
\newcommand{\rb}{\right)}
\newcommand{\lcd}{\Lambda {\rm CDM}}
\def\LCDM{$\Lambda$CDM}

\usepackage{tikz}
\usepackage{cancel}

\begin{document}

\title{Studying the Peculiar Velocity Bulk Flow in a Sparse Survey of Type-Ia SNe}

\author[B. Rathaus, E. D. Kovetz and N. Itzhaki]{Ben Rathaus$^{1}$, Ely D. Kovetz$^{2}$ and Nissan Itzhaki$^{1}$ \\
$^1$Raymond and Beverly Sackler Faculty of Exact Sciences,\\
School of Physics and Astronomy, Tel-Aviv University, Ramat-Aviv, 69978, Israel\\
$^2$Theory Group, Department of Physics and Texas Cosmology Center, \\
The University of Texas at Austin, TX, 78712, USA}

\date{E-mail: ben.rathaus@gmail.com\\ E-mail: elykovetz@gmail.com \\ \\ In original form: \today{}} 
\pagerange{\pageref{firstpage}--\pageref{lastpage}}
\maketitle\label{firstpage}
\begin{abstract}
Studies of the peculiar velocity bulk flow based on different tools and datasets have been consistent so far in their estimation of the direction of the flow, which also happens to lie in close proximity to several features identified in the cosmic microwave background, providing motivation to use new compilations of type-Ia supernovae measurements to pinpoint it with better accuracy and up to higher redshift.
Unfortunately, the peculiar velocity field estimated from the most recent Union2.1 compilation suffers from large individual errors, poor sky coverage and low redshift-volume density. We show that as a result, any naive attempt to calculate the best-fit bulk flow and its significance will be severely biased.
Instead, we introduce an iterative method which calculates the amplitude and the scatter of the direction of the best-fit bulk flow as deviants are successively removed and take into account the sparsity of the data when estimating the significance of the result.
Using 200 supernovae up to a redshift of z=0.2, we find that while the amplitude of the bulk flow is marginally consistent with the value expected in a $\Lambda {\rm CDM}$ universe given the large bias, the scatter of the direction
is significantly low (at $\gtrsim99.5\%$ C.L.) when compared to random simulations, supporting the quest for a cosmological origin.

\end{abstract}

\begin{keywords}
peculiar velocities -- bulk flow -- supernovae: type-Ia
\end{keywords}

\section{Introduction}
In the last couple of decades a considerable effort has been devoted to the analysis of the peculiar velocity field in search for an overall bulk flow (BF) on ever increasing scales, lately reaching as high as $\sim\!100\,{\rm Mpc}/h$ using galaxy surveys \citep{Dressler:1986rv, LyndenBell:1988qs, Gorski:1988, Courteau:2001ve, Hudson:2004et, Sarkar:2006gh, Watkins:2008hf, Sheth:2008ef, Feldman:2009es, Ma:2010ps, Macaulay:2010ji, Nusser:2011tu, Nusser:2011sd, Branchini:2012rb, Ma:2012wp, Ma:2012tt} 
and type-Ia supernovae (SNe) \citep{Colin:2010ds, Haugboelle, Davis:2010jq, Turnbull:2011ty, Dai:2011xm, Riess:1995cg, Riess:1997ai}  and even an order of magnitude higher, based on measurements of the kinetic Sunyaev-Zeldovich effect in the cosmic microwave background (CMB) \citep{Kashlinsky:2008ut, Keisler:2009nw, Mak:2011sw, Kashlinsky:2012gy, Mody,Lavaux}. While there have been conflicting claims regarding the amplitude of the dipole moment of this field and its tension with the expected value in a \LCDM{} universe, the vast majority of these surveys have been consistent in their findings for the direction of the dipole\footnote{Using type-Ia SNe, \citep{Kalus:2012zu} recently found that the direction of highest cosmic expansion rate is also in the vicinity of this dipole, though it is consistent with the expectation from \LCDM{}.}.

Meanwhile, several features in the CMB temperature maps from the COBE DMR \citep{COBE} and WMAP \citep{WMAP} experiments have been identified in roughly the same region of the sky, from the dipole moment \citep{Smoot} to several reported anomalies, including the alignment between the quadrupole and octupole \citep{Tegmark,Land:2005ad}, mirror parity \citep{Land:2005jq, BenDavid:2011fc, Finelli} and giant rings \citep{Kovetz:2010kv}. This coincidence provides further motivation to search for a unified cosmological explanation \cite{Perivolaropoulos:2011hp}.

Over the years a number of cosmological scenarios have been suggested as possible sources for a peculiar velocity BF, such as a great attractor \citep{Dressler:1986rv, LyndenBell:1988qs}, a super-horizon tilt \citep{TurnerTilt}, over-dense regions resulting from bubble collisions \citep{LarjoLevi} or induced by cosmic defects\footnote{An over-density induced by a pre-inflationary particle would imprint giant  rings in the CMB whose center is aligned with the BF \citep{Fialkov}.} \citep{Fialkov,DomainWall}, etc. In an attempt to test these hypotheses and distinguish between them, any knowledge regarding the redshift dependence of the BF can be a crucial discriminator.

Type-Ia SNe, whose simple scaling relations provide empirical distance measurements and which have been detected up to redshifts $z\gtrsim1$, provide a unique tool to estimate the peculiar velocity BF and study its direction and redshift extent.
However, this approach also contains certain caveats. First, datasets from typical type-Ia SNe surveys are orders of magnitude smaller than those from galaxy surveys and their sky coverage and redshift-volume density are extremely poor. Secondly, different SNe compilations often use different light-curve fitters, involving different nuisance parameters. Currently, the most promising candidate for a large scale BF search is the Union2.1 compilation \citep{union21} (see also \citep{Kowalski:2008ez, Amanullah:2010vv}), comprising of 19 different surveys which are all analyzed with a single light-curve fitter (SALT2 \citep{SALT2})

The purpose of this work is to investigate the peculiar velocity field extracted from the Union2.1 data and given its limitations determine which conclusions can be reliably made as to the BF in the inferred radial peculiar velocity field, placing an emphasis on its direction and redshift extent. Accounting for the substantial bias due to the sparsity of the data and using a dedicated algorithm to iteratively remove outlying data points from the analysis, we test the amplitude and the scatter of the direction of the BF and estimate the significance of the results using Monte Carlo simulations.

The paper is organized as follows. In Section 2 we describe the initial filtering of the data and the method for extracting the individual radial components of the peculiar velocities, as well as how we generate random simulations of data with the same spatial distribution. In Section 3 we address the inevitable bias due to sparsity in both the amplitude and direction in naive best-fit methods used to detect an overall BF. We introduce our method in Section 4 and define a score which measures the scatter of the best-fit direction in successive iterations. We demonstrate that this score is effective in identifying simulated datasets with an artificially inserted BF and discuss how the significance of its findings can be estimated. In Section 5 we consider both the full dataset and the application of the scatter-based iterative method to the data and present the results. We conclude in Section 6.

\section{Preliminaries}
\subsection{Data filtering}
We use the recent type-Ia SNe compilation Union2.1 \citep{union21}, which contains 580 filtered SNe at redshifts $ 0.015\!<\!z\!<\!1.4 $. This compilation is drawn from 19 datasets, all uniformly analyzed with a single light-curve fitter (SALT2 \citep{SALT2}), and analyzed in the CMB-frame. At high redshifts, the spatial distribution of this dataset grows increasingly sparse, the individual errors become large and some of the induced radial peculiar velocities, calculated as described below, take on unreasonable values (such as $>\!0.5{\rm c}$). In order to avoid these pathologies while still retaining the ability to examine the behavior at distances larger than those accessible with galaxy surveys ($\lesssim100\,{\rm Mpc}/h$), we apply an initial cutoff in redshift and remove all points with $z>0.2$ (corresponding to $\lesssim550\,{\rm Mpc}/h$) from our dataset. 

\begin{figure}
\centering
\includegraphics[width=0.5\textwidth]{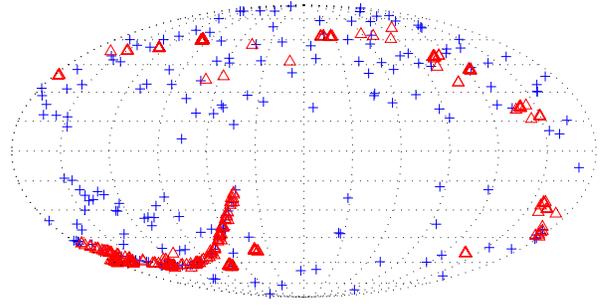}
\caption{The spatial scatter of all SNe in the Union2.1 compilation. The red triangles indicate SNe with $z>0.2$ that are filtered out before any analysis is performed.}
\label{fig:dataSpatialScatter}
\end{figure}

In Fig.~\ref{fig:dataSpatialScatter} we plot the spatial distribution of the 
Union2.1 dataset with the remaining $200$ SNe marked in blue.
It is apparent that even outside the galactic plane the sky coverage is quite poor and the three-dimensional distribution of the remaining data is significantly sparse and inhomogeneous. The implications of this will be discussed in the next section.

\subsection{Peculiar velocities and best-fit bulk flow}
The Union2.1 dataset specifies for each SN its measured redshift $z$, the inferred distance modulus $\mu_{\text{obs}}$ and the error $\Delta\mu_{\text{obs}}$. The relation between the cosmological (``{\it true}'') redshift $\bar{z}$ and the distance modulus is given by
\be \label{eq:disMod}
\mu = 5\log_{10}\lb \frac{d_L(\bar{z})}{1~\text{Mpc}}\rb   + 25 ,
\ee 
where $d_L$ is the luminosity distance, which in a flat universe with matter density $\Omega_M$, a cosmological constant $\Omega_{\Lambda}$ and a current Hubble parameter $H_0$, is
\be\label{eq:dL}
d_L (\bar{z})= \frac{(1+\bar{z})}{H_0} \int_{0}^{\bar{z}} \frac{\mathrm{d}z' } {\sqrt{ \Omega_m(1+z')^3 + \Omega_{\Lambda}}}.
\ee 
Due to the peculiar velocity, both the observed redshift and distance modulus (through the luminosity distance) will differ from their {\it true} cosmological values \citep{Hui:2005nm}. To first order in $\mathbf{v}\cdot \mathbf{\hat{n}}$, where $\hat{\mathbf{n}}$ is the direction pointing to a SN with peculiar velocity ${\bf v}$, we get
\ben
~~~~~~~~~~~~~~~~~~1+z &=& (1+\bar{z})(1+\mathbf{v}\cdot \mathbf{\hat{n}}) \nonumber\\
~~~~~~~~~~~~~~~~~~d_L(z)&=&d_L(\bar{z})(1+2\,\mathbf{v}\cdot \mathbf{\hat{n}}).
\een
In order to extract the radial peculiar velocity $v_r$ of the SNe in our dataset, we follow the first order expansion in \citep{Haugboelle, Hui:2005nm} 
\be \label{eq:vr}
v_r=-\frac{\ln10}{5}\frac{H(z)d_A(z)}{1-H(z)d_A(z)}(\mu_{\text{obs}}-\mu(z)),
\ee
where $H(z)$ is the Hubble parameter at redshift $z$, and $d_A(z) = d_L(z)/(1+z)^2$ is the observed angular diameter distance to the SN. 

We then find the best-fit BF velocity $\mathbf{v}_{\rm {BF}}$ in our set of $N$ SNe, each with a radial velocity amplitude $v_{r}^{i}$ in a direction $\hat{\mathbf{n}}_i$, by minimizing 
\be
\chi^2(\mathbf{v}_{\rm BF}) = \sum_i \frac{(v_{r}^{i}
-\mathbf{v}_{\rm BF}\cdot\hat{\mathbf{n}}_i)^2}{(\Delta v_{r}^{i})^2}
\label{eq:bfbf}
\ee
with respect to the direction and amplitude of $\mathbf{v}_ {\rm {BF}}$, where $\Delta v_{r}^{i}$ are the individual errors obtained from the measurement errors in the distance moduli $\Delta \mu_{\text{obs}}^{i}$ using Eq.~(\ref{eq:vr}).

\subsection{Monte Carlo simulations}
\label{sec:MC}
Our Monte Carlo simulations consist of random permutations of the sky locations 
of the SNe in our dataset, after removing the initial BF velocity 
$\mathbf{v}_{\rm {BF_{\rm init}}}$ from the entire set by subtracting 
its corresponding component from the individual velocities 
\be \label{eq:vsubt}
v_{r}^{i} \longrightarrow v_{r}^{i} - \mathbf{v}_{\rm {BF_{\rm init}}}
\cdot \hat{\mathbf{n}}_{i}. 
\ee
The new dataset will have the same spatial distribution and its own 
initial random BF with a typical $v_{\rm rms}$ amplitude. In order 
to simulate a random realization with a specifically chosen cosmological BF (up to statistical noise), for the purposes of testing our method, we simply add the chosen non-random BF contribution to the individual velocities after the permutation.

For the analysis in this paper we use $16,000$ random realizations 
(spatial permutations) of the Union2.1 data with no artificially inserted BF in order to test against the null hypothesis. To examine the detection capabilities of our method, we use $6,000$ different random realizations for each inserted BF amplitude in the range $|{\bf v}_{\rm BF}|=\{50,100\dots450~\rm{km/s}\}$, all in the direction $(l,b)=(295^\circ,5^\circ)$, which is the direction of the best-fit BF on the full dataset (for the purposes of estimating the significance of our results, we have verified that this specific choice of direction has no effect). 

\section{Sparsity Bias}
As mentioned above, the spatial distribution of the SNe 
dataset is inhomogeneous and sparse across the sky and in redshift depth. 
As a consequence, any search for an overall BF will be severely biased. 
Such a bias must be taken into account when evaluating the significance 
of a measured best-fit BF vs. the expectation from a \LCDM{} universe.
We now examine this bias separately in terms of the direction and amplitude of the BF.
In the first subsection we show that the sparsity of our dataset causes a preference for a flow in directions within the galactic plane. In the second subsection we show that the root-mean-square (rms) velocity typically used under the \LCDM{} hypothesis is inappropriate for a significance estimation of the BF amplitude in a sparse dataset such as ours.

\subsection{Bulk flow - direction} 
\label{sec:BFDirection}
In a dense homogeneous dataset (which has no preferred direction), if we perform many random mixings of the sky locations of the SNe, the best-fit BF direction will be distributed uniformly over the $4\pi$ area of the sky.
In a histogram of the measured directions, inside a circle of radius $\alpha$ around any sky coordinate we expect to find a fraction 
\be
f(\alpha) = \sin^2(\alpha/2)
\ee
of the results. 

\begin{figure*}
\centering
\includegraphics[width=0.48\linewidth]{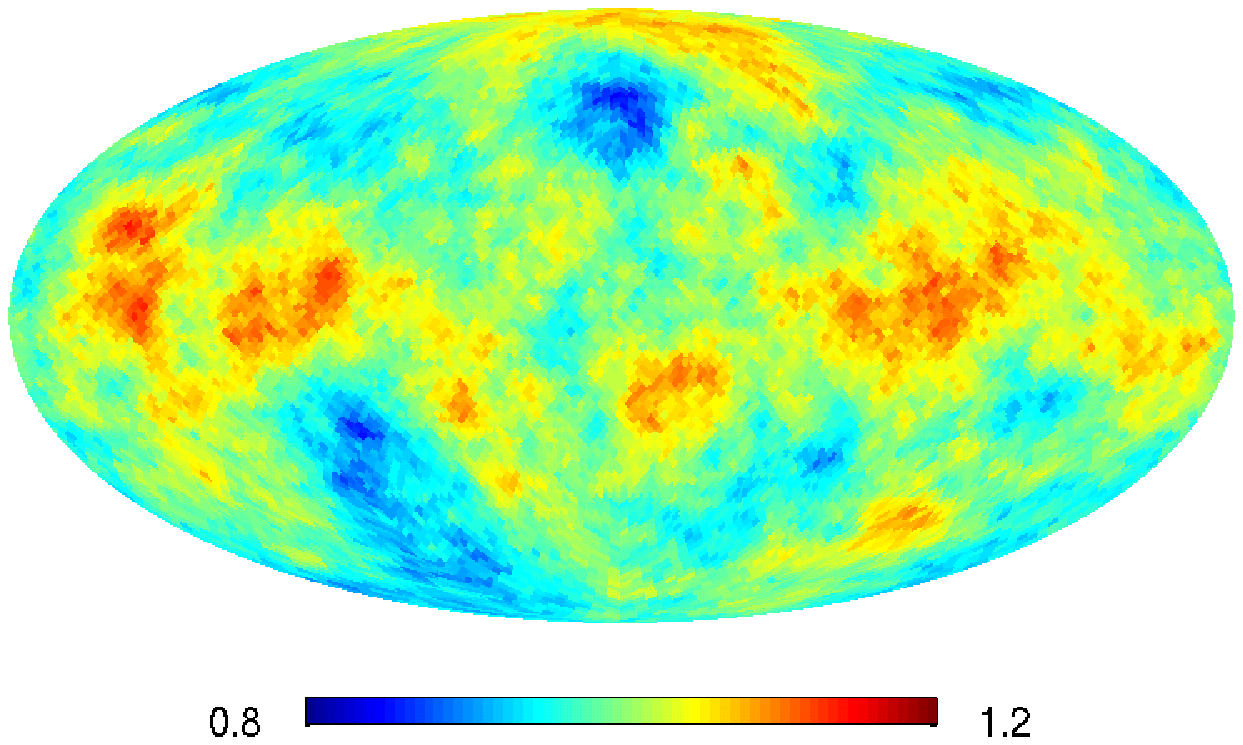}
\includegraphics[width=0.48\linewidth]{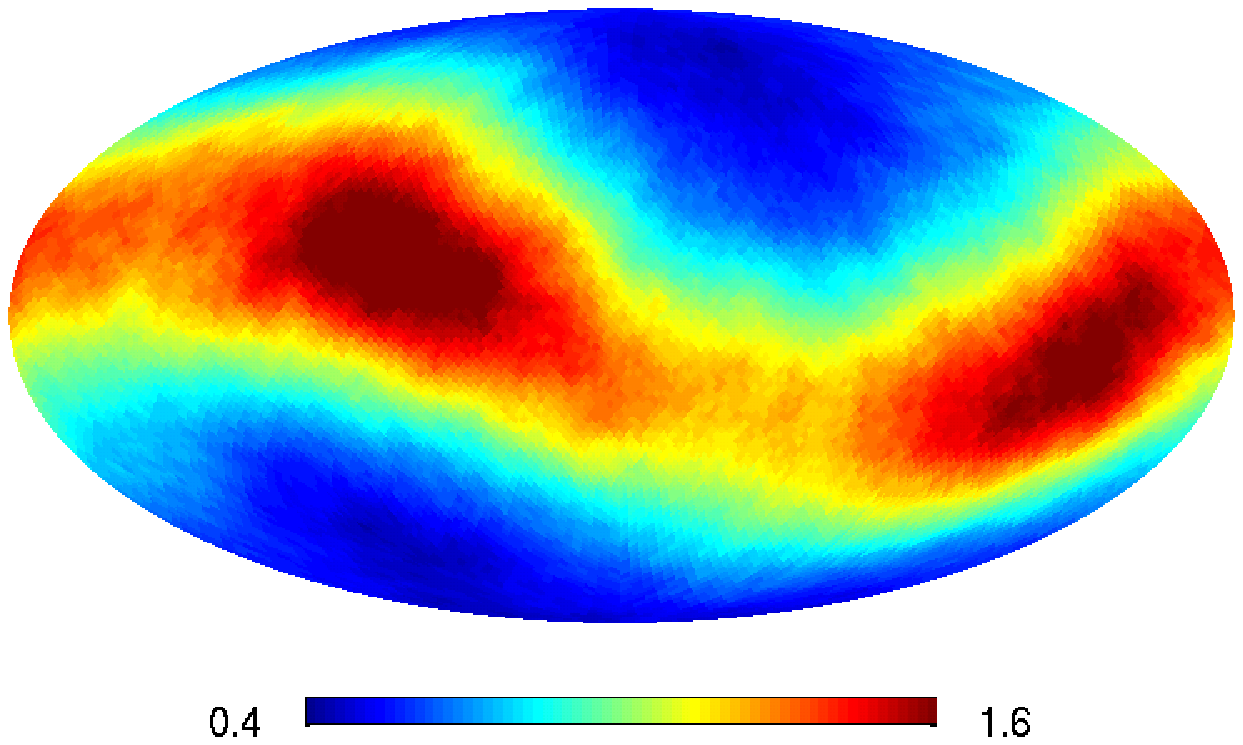}
\label{homogF}
\caption{{\it Left:} The ratio between the measured and expected fractions $f_{\rm{meas.}}/f$ of $16,000$ randomly picked directions (simulating the homogeneous case) that point within $\alpha=20^\circ$ from each coordinate. {\it Right:} The same ratio inferred from $16,000$ random permutations of the locations of SNe of Union2.1 dataset, where the preference for the galactic plane is clearly seen.}
\label{fig:homogeneity}
\end{figure*}

To demonstrate the bias induced by the sparsity of our dataset, we plot in 
Fig.~\ref{fig:homogeneity} the ratios between the measured fraction  
and its expected value $f_{\rm{meas.}}/f$ for a uniformly distributed set ({\it Left}) as well as for our dataset ({\it Right}), using $\alpha=20 ^\circ$.
We see that the result for our dataset is far from isotropic. Its spatial distribution, regardless of the observed magnitudes, is biased towards a specific portion of the sky, namely the region surrounding 
the galactic plane.  

\subsection{Bulk flow - amplitude} 
\label{sec:BFAmplitude}
Another implication of the sparsity in the data is that the random component 
of the BF does not follow the expected \LCDM{} behavior. 
We must therefore quantify the difference between the expected \LCDM{} rms velocity and the rms velocity we expect when dealing with a sparse dataset such as Union2.1. 

\subsubsection{Velocity rms in \LCDM{}}
In \LCDM{}, the expected value for the BF amplitude is zero 
$\left\langle\mathbf{v}\right\rangle = 0$ while its variance satisfies \citep{Gorski:1988}
\be\label{eq:LCDMrms}
\sigma_{\Lambda}^{2}
\equiv \left\langle \mathbf{v} \cdot \mathbf{v} \right\rangle = \frac{H_0^2f^2}{2\pi}\int \mathrm{d}k P(k)\left|W(kR)\right|^2,
\ee
where $f=\Omega_{m}^{0.55}$ is the dimensionless linear growth rate, 
$P(k)$ is the matter power spectrum, $W(kR)$ is the Fourier transform 
of a window function with characteristic scale $R$ and the angle brackets 
$\left\langle .. \right\rangle$ denote an ensemble average. Since \LCDM{} is 
isotropic, this means that for each primary direction $i\in\{x,y,z\}$ in a Cartesian 
coordinate system the BF amplitude may be described using a normal distribution 
\be\label{eq:normDist}
v_i \sim \mathcal{N} (0, \sigma_{\Lambda}/\sqrt{3}), \quad i=x,y,z.
\ee
To estimate the significance of a non-vanishing BF measured in a given 
survey, a common approach is to tweak the frame of reference so that the BF 
points exactly in one of the primary directions, e.g. $\hat{\mathbf{e}}_y$, and 
compare the ``single component'' measured BF to $\sigma_{\Lambda}/\sqrt{3}$.
However, this ignores the fact that the BF amplitude in the other two directions vanishes due to this particular choice of frame and 
would lead to an overestimated significance of the BF amplitude.

To resolve this, we use the fact that the BF amplitude is a square-root of a sum of three normally distributed variables $|{\bf v}|^2=\left( \sum_i v_{i}^{2}\right)$, and so in \LCDM{} it satisfies
\be\label{eq:chiDist}
\lcd\text{: } |{\bf v}| \sim \chi_3(\sqrt{3}x/\sigma_\Lambda).
\ee
That is, it follows an ``unnormalized'' $\chi$ distribution with $3$ degrees of freedom (a Maxwell-Boltzmann distribution) \citep{Li:2012cv}. Eq.~(\ref{eq:chiDist}) represents the probability density 
function (PDF) of the BF amplitude inside some volume, that is modulated 
by the same window function as in Eq.~(\ref{eq:LCDMrms}), in an unbiased way.

\subsubsection{Velocity rms in Union2.1}

In order for the right-hand side of Eq.~(\ref{eq:chiDist}) to describe the observed BF $|{\bf v^{\rm obs}}|$  appropriately, one needs to measure the peculiar velocity in many spatial locations, so that the typical separation between any two nearest neighbors that were measured will be much smaller than the coherence scale.
This is clearly not satisfied for the sparse Union2.1 dataset. Therefore we should replace the window function with a sum of $N$ delta functions, each centered on the location $\mathbf{R}_i$ of a single SN
\be\label{eq:windowToDelta}
W(\mathbf{r}) \rightarrow \frac{1}{N}\sum_{i=1}^{N} \delta(\mathbf{r}-\mathbf{R}_i).
\ee
However, since in Fourier space 
\be
\delta(\mathbf{r} -\mathbf{R}_i) \rightarrow \exp\{-i \mathbf{k}\cdot \mathbf{R}_i \},
\ee
the new window function term will consist of $\sim N^2$ interference terms that are no longer spherically symmetric. Therefore using Eq.~(\ref{eq:windowToDelta}) to evaluate the sparse-case equivalent of Eq.~(\ref{eq:chiDist}) is unfeasible. Instead we use the amplitudes of the best-fit BF of 
$16,000$ random spatial permutations of our dataset, as described in \S\ref{sec:MC}, as an approximation of the PDF for the BF amplitude inside a sphere of radius $z=0.2$. 
The difference between this approximation and the isotropic \LCDM{} 
scenario will be encoded in a best-fit $\sigma_{\rm{fit}}$ (instead of $\sigma_{\Lambda}$) which describes the observed distribution 
\be\label{eq:chiDistRand}
\text{Sparsity: } |{\bf v^{\rm obs}}| \sim \chi_3(\sqrt{3}x/\sigma_ {\rm{fit}}).
\ee

\begin{figure}
\centering
\includegraphics[width=0.5\textwidth]{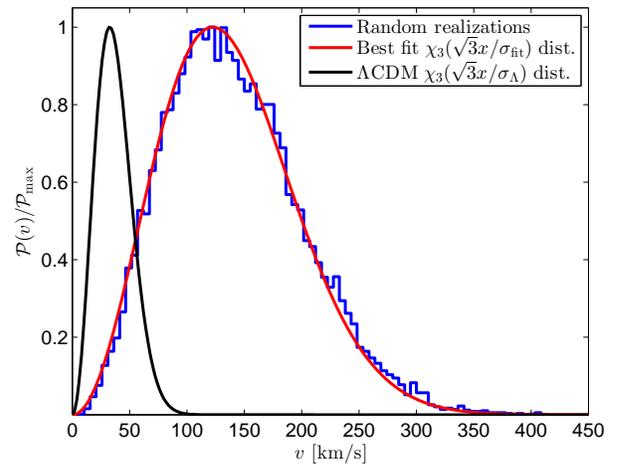}
\caption{The normalized PDF for the amplitude of the BF for \LCDM{} inside a sphere of radius $550~\rm{Mpc/} h$, calculated according to Eqs.~(\ref{eq:normDist})-(\ref{eq:chiDist}) ({\it black}) as well as for random realizations given the sparsity of the Union2.1 data ({\it blue}). The red line is the best fit $\chi_3(\sqrt{3}x/\sigma_{\rm{fit}})$ according to
$16,000$ random permutations of the locations of our dataset, as described in \S\ref{sec:BFAmplitude}. The goodness of fit is $R^2=0.9945$.}
\label{fig:vprob}
\end{figure}

In Fig.~\ref{fig:vprob} we plot the approximated PDF for the BF amplitude and the corresponding best-fit  
$\chi_3$ distribution according to Eq.~(\ref{eq:chiDistRand}). We find 
\be\label{eq:Sparserms} 
\sigma_{\rm{fit}} \approx 150~\text{km/s}
\ee
as opposed to $\sigma_\Lambda = 43~\text{km/s}$ calculated 
directly from Eq.~(\ref{eq:LCDMrms}) for a top-hat window function of size $R=550~\rm{Mpc/}h$. 
We see from Fig.~\ref{fig:vprob} that a naive estimation of the significance of a measured BF amplitude in a sparse survey would be highly overestimated.

\section{Scatter-Based Method}
\label{sec:SBMethod}
We present a method based on an iterative process of repeatedly fitting a BF to the peculiar velocity field after the removal of the datapoint with the highest deviation from the previous fit. If there is a significant BF in the full dataset, the compactness of the scatter in the directions identified for the best-fit BF in each iteration can be used as an efficient estimator of the significance of the original flow. The stronger the flow in the full dataset, the smaller the scatter we will measure in the iterations.

\subsection{Iterative algorithm} 
Heuristically, the algorithm can be sketched as follows
\begin{center}
\begin{tikzpicture}[bend angle=45]
\node  (critical) at (0,0){$\rightarrow$ residuals$ \rightarrow$};
\node (ol) at (1.7,0) {outlier};
\node (bf) at (-1.9,0) {BF fitting};
\path[->] (ol.east) edge [out=340, in=200] node [below right=0.3] {iterations} (bf.west);
\end{tikzpicture}
\end{center}
After calculating the best-fit BF of the complete set according to Eq.~(\ref{eq:bfbf}), we examine the residual velocities of the different SNe in order to identify the one with the strongest deviation from the bulk.
In each iteration, we find the best-fit $\mathbf{v}_{\rm BF}^{\rm iter}$ and then identify the point $i$ with the largest contribution 
\be
\Delta\chi^2(\mathbf{v}_{\rm BF}^{\rm iter})=(v_{r}^{i}-\mathbf{v}_{\rm BF}^{\rm iter}\cdot \hat{\mathbf{n}}^i)^2/ (\Delta v_{r}^{i})^2
\ee
and remove it from the dataset before the next iteration. By iteratively removing these deviants, we can also verify that our results are not dominated by a small subset of the data with some common characteristic such as low redshift or a specific location on the sky.

\subsection{Scatter score}
To measure the scatter, we assign a {\it scatter score} to the data, defined by
\be
S = \sum_{j=2}^{N_{\rm iter}} \arccos(\hat{\mathbf{n}}_j\cdot \hat{\mathbf{n}}_1) + \arccos(\hat{\mathbf{n}}_j\cdot \hat{\mathbf{n}}_{j-1}),
\ee
which is a cumulative sum of the consecutive and total shifts, i.e. the sum of the distances from the direction of the best-fit BF in the current iteration $\hat{\mathbf{n}}_j$ to the one in the last iteration $\hat{\mathbf{n}}_{j-1}$ and to that of the first iteration $\hat{\mathbf{n}}_1$. This measures both the tightness and the extent of the scatter of the measured directions throughout the iterative process.

\begin{figure}
\centering
\includegraphics[width=0.5\textwidth]{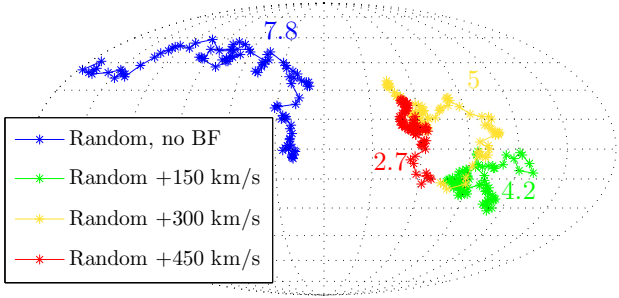}
\caption{Illustration of the scatter of the best-fit BF direction in each iteration for a random realization with no inserted flow ({\it blue}) and for various sets with artificially inserted BF amplitudes, all in the direction $(l,b)=(295^\circ,5^\circ)$. The total number of iterations shown here is $191$, after which only $10$ SNe were left in the dataset. The score for each realization is shown in the appropriate color, normalized by the score of the real data.}
\label{fig:scatterRandom}
\end{figure}

In Fig.~\ref{fig:scatterRandom} we demonstrate the results for the scatter score $S$ by plotting the directions of the best-fit BF at each iteration for a random realization with just a random BF and with increasing artificially-added BF amplitudes in the direction $(l,b)=(295^\circ,5^\circ)$, which is the direction of the best-fit BF of our dataset (this choice of inserted direction allows a straightforward comparison with the data and accounts for a possible bias, as mentioned in \S\ref{sec:BFDirection}, but we have verified that it has no effect on our significance estimation).
The results are consistent with the expected behavior: as the inserted BF amplitude is increased, the scatter becomes smaller and converges to a region closer to the inserted BF direction. 

\subsection{Significance estimation}
\label{sec:significance}
We compare $S_{\text{data}}$ with the mean value $S$ evaluated using $6,000$ random simulations for each inserted BF amplitude $|{\bf v_{\rm BF}}|$ and infer the significance of the data in terms of the probability that a random \LCDM{} realization will get a lower score than the data
\be\label{eq:prob}
\mathcal{P}(S < S_{\text{data}})=\int{ \mathcal{P}(S<S_{\text{data}} \left| \right. |{\bf v}|) \mathcal{P}(|{\bf v}|)} \mathrm{d} |{\bf v}|,
\ee
where $\mathcal{P}(|{\bf v}|)\mathrm{d}|{\bf v}|$ is the probability that a \LCDM{} realization of the data will have a BF of amplitude between $|{\bf v}|$ and $|{\bf v}|+\mathrm{d}|{\bf v}|$ given the sparsity of the data according to \S\ref{sec:BFAmplitude}, and $\mathcal{P}(S<S_{\text{data}} \left| \right. |{\bf v}|)$ is the conditional probability that a random simulation will result in $S<S_{\text{data}} $ given a BF amplitude $|{\bf v}|$.

\vspace{-0.05in}
\section{Results}

\subsection{Full dataset}
Before applying our scatter-based method described in the last section, we note that using a naive best-fit, the overall BF in our dataset has an amplitude $|{\bf v_{\rm BF}}|=260~\text{km/s}$ and points in the direction $(l,b)=(295^\circ,5^\circ)$, which is in agreement with results reported elsewhere \cite{Hudson:2004et, Sarkar:2006gh, Watkins:2008hf, Feldman:2009es, Ma:2010ps, Riess:1995cg, Colin:2010ds, Haugboelle, Davis:2010jq, Dai:2011xm, Turnbull:2011ty, Mak:2011sw, Kashlinsky:2012gy, Lavaux}. This direction lies in proximity to features in the CMB (most of all to the giant rings reported in \cite{Kovetz:2010kv}), but is also close to the galactic plane, as might have been expected given the sparsity bias shown in Fig.~\ref{fig:homogeneity}. 
In addition, referring back to Fig.~\ref{fig:vprob} we see that when comparing with the expected rms amplitude in a finite-size survey with the same spatial distribution, this amplitude, although high, is consistent with \LCDM{} at the $95\%~$C.L. (naively using the unbiased \LCDM{} expectation, one might have assigned a much larger significance to this result).

Thus, using the full dataset, we conclude that no claim can be made as to the existence of a cosmological BF in the Union2.1 type-Ia SNe data up to redshift $z=0.2$ given the significant bias induced by the poor sky coverage and redshift-volume density of this dataset. 
\begin{figure*}
\centering
\includegraphics[width=0.48\textwidth]{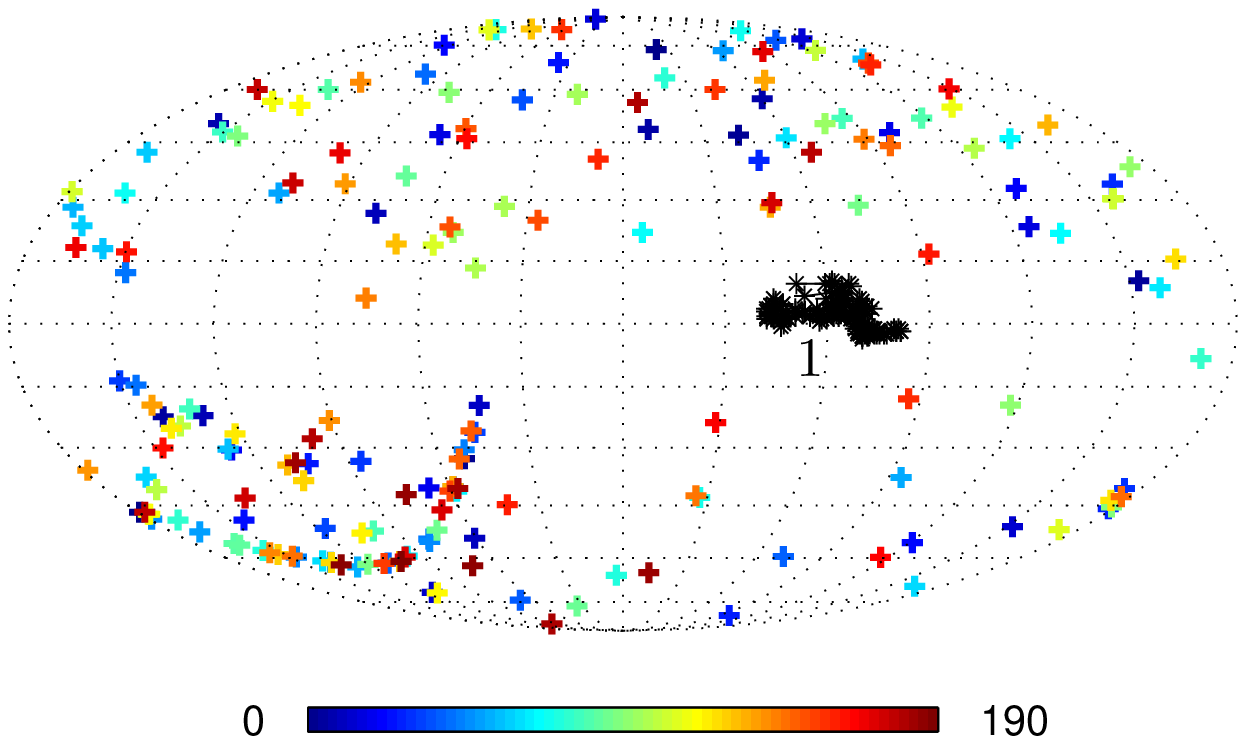}
\includegraphics[width=0.42\textwidth]{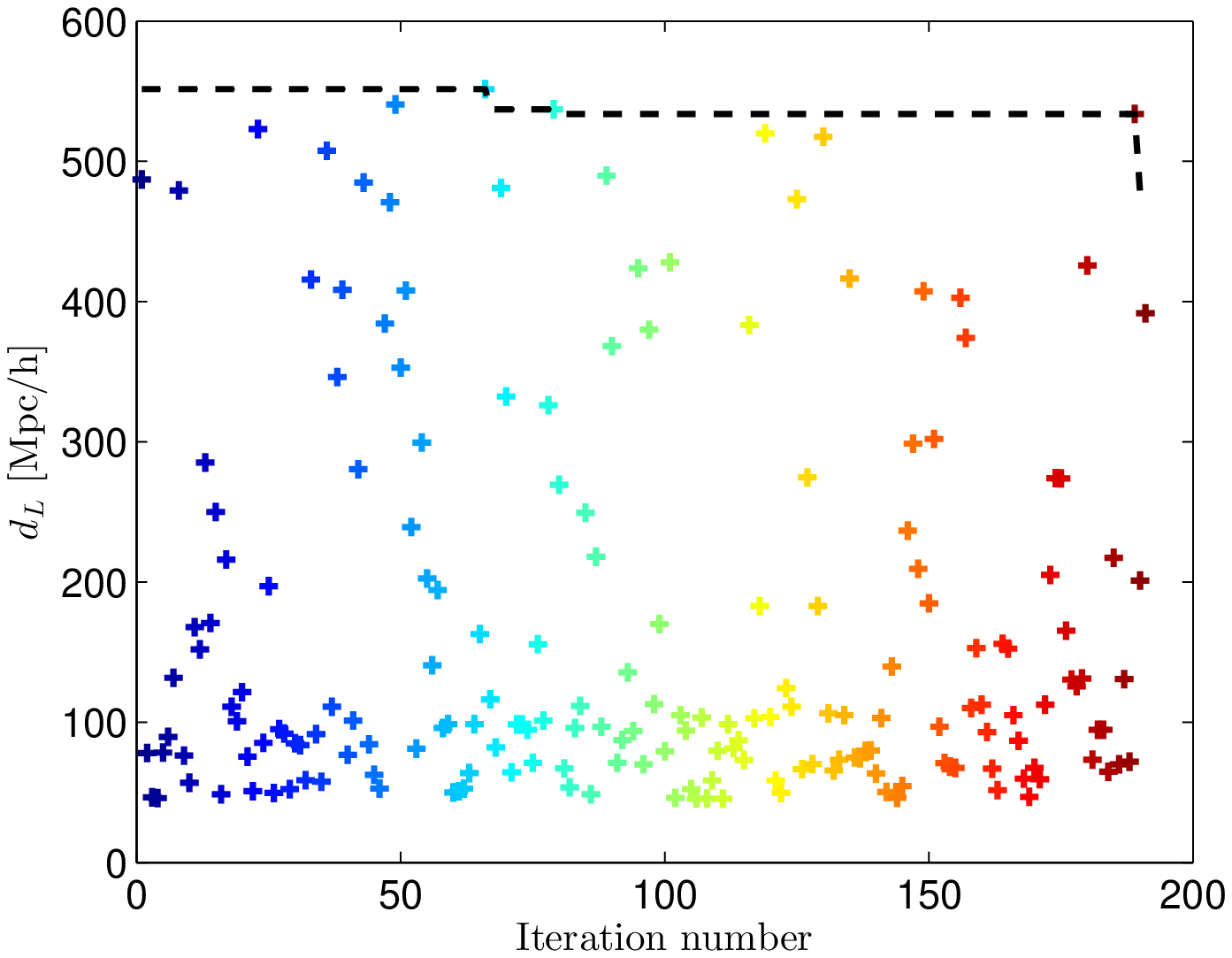}
\caption{{\it Left:} The scatter of the best-fit BF in the Union2.1 dataset at redshifts $z<0.2$. The black stars mark the best-fit BF direction of each iteration until only 10 SNe are left in the dataset. The plus signs indicate the spatial coordinate of the SN that is excluded at each iteration and are coloured as a function of the iteration number ({\it blue} - excluded first, {\it red} - excluded last). {\it Right:} The distance to the excluded SN in each iteration as a function of the iteration number. The dashed line marks the distance to the farthest SN remaining in the dataset at each iteration.}
\label{fig:scatter}
\end{figure*} 

\subsection{Sifting iteratively through the data}
\label{sec:fiftyP}
The scatter of the best-fit BF direction measured in the iterative process described in \S\ref{sec:SBMethod} is plotted in the left panel of Fig.~\ref{fig:scatter}. In the right panel we plot the luminosity distance to the excluded SN as a function of the iteration number, and demonstrate that our results are not dominated by a subset of nearby SNe (SNe at distances $\gtrsim500\,{\rm Mpc}/h$ remain in the dataset until the final iterations).
Comparing Figs.~\ref{fig:scatterRandom}  ({\it Left}) and \ref{fig:scatter}  ({\it Left}) we see that the scatter of the data is much smaller than that of the realizations with $|{\bf v}_{\rm BF}|\leq300~\text{km/s}$, and is comparable in size to a $|{\bf v}_{\rm BF}|\gtrsim450~\text{km/s}$ realization. 

\begin{figure*}
\centering
\includegraphics[width=\textwidth]{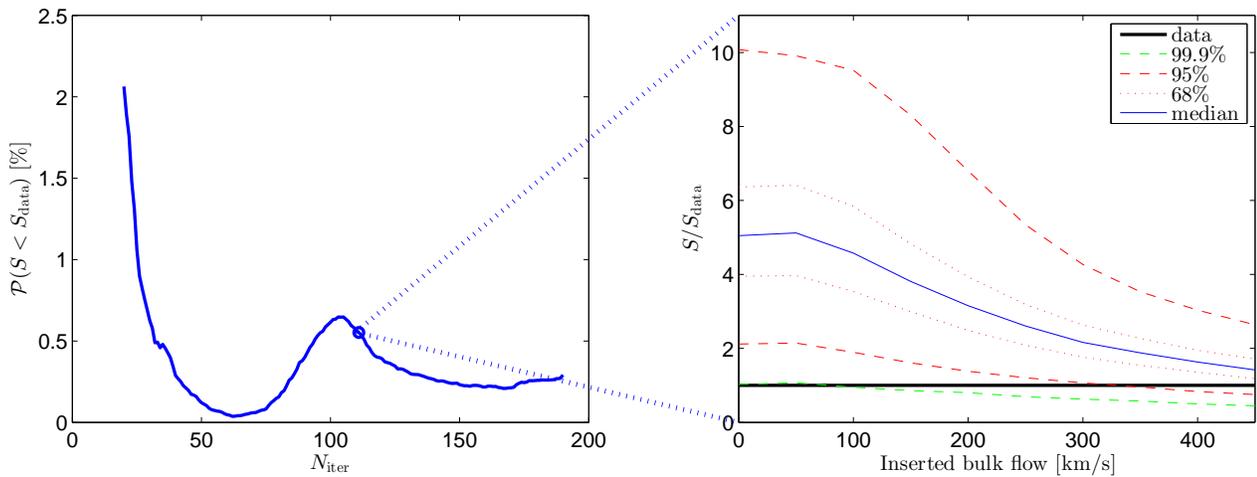}
\caption{{\it Left:} The significance of the compactness of the scatter with respect to \LCDM{}, as described by Eq.~(\ref{eq:prob}), as a function of the total number of iterations $N_{\rm{iter}}$, {\it Right:}  A few percentiles of the results of the normalized score evaluated for random realizations of the data according to (\ref{eq:prob}), along with the data result, for $N_{\rm{iter}}\!=\!110$.}
\label{fig:new}
\end{figure*}

This is also shown quantitatively in Fig.~\ref{fig:new} ({\it Left}), where we plot the significance of the compactness of the scatter with respect to \LCDM{}, as described by Eq.~(\ref{eq:prob}), as a function of the total number of iterations $N_{\rm{iter}}$. For $N_{\rm{iter}}\!=\!110$ (chosen arbitrarily), we show in Fig.~\ref{fig:new} ({\it Right}) a few percentiles of the results of the normalized score evaluated for random realizations of the data according to (\ref{eq:prob}), along with the data result. We see that the score for our dataset is outside the $95\%$ C.L. for any initial (i.e. for the whole dataset) BF amplitude smaller than $300\,{\rm km/sec}$. 

Integrating over all possible initial BF amplitudes, we see that $S_{\text{data}}$ is surprisingly low with respect to the expectation from a \LCDM{} universe: the overall probability that a single \LCDM{} realization would get a score that is as low as the score of the real data is $<\!0.5\%$ for any $N_{\rm{iter}}\!>\!30$, and gets as low as $0.1\%$ for some choices of $N_{\rm{iter}}$. Thus we conclude that the scatter of the best fit BF direction is significantly low, at a $\gtrsim 99.5\%$ C.L.

\section{Conclusions}
The goal of this work was to use the most recent compilation of type-Ia SNe measurements in order to test the claims of a peculiar velocity BF in different studies. After truncating the Union2.1 catalogue at redshift $z\!=\!0.2$ and extracting the radial peculiar velocity field, we showed that a naive attempt to measure a best-fit BF in this field ignores a significant bias due to its sparse spatial distribution and renders inconclusive results for the amplitude and direction of the best-fit flow.
This sparsity bias was discussed in detail above along with the difficulty in determining the correct \LCDM{} prediction with which any result should be compared. We presented a prescription for estimating this value in a finite survey of given redshift extent and spatial distribution, and concluded that the BF amplitude measured in the Union2.1 data up to redshift $z\!=\!2$ is consistent with the $95\%$ C.L. limits.

Given the consistency in the reports from a wide spectrum of analyses regarding the direction of the measured BF and the alignment between the reported values and certain CMB features, we focused on the direction and introduced a method which measures the scatter in the best-fit BF direction as outlying points are removed in iterations. We were careful to use realistic expectations for a BF amplitude in a sparse dataset and used Monte Carlo simulations with similar sparsity to estimate the significance of our findings. Our results suggest that the Union2.1 data up to redshift $z=0.2$ contains an anomalous BF at a $99.5\%$ C.L. compared to random simulations with the same sparsity as the data.
   
In the future, as more data is collected, the method used in this work will become more and more robust and enable the measurement of the BF in consecutive redshift bins to yield a better analysis of the redshift dependence of the measured result. In addition, it might be possible to focus on measurements from a single survey and thus reduce the errors stemming from combining several surveys with different characteristics.

If the reports of a BF which is inconsistent with \LCDM{} are verified by future observations, it shall serve as a promising lead for theoretical research exploring areas beyond the concordance cosmological model. The full potential of type-Ia SNe data to settle this issue is yet to be realized.
\vspace{-0.1in}
\section*{Acknowledgments}

BR thanks A. Nusser and D. Poznanski for useful discussions. We also thank A. Nusser for comments on the first version of this work. EDK was supported by the National Science Foundation under Grant Number PHY-0969020 and by the Texas Cosmology Center.

\label{lastpage}

\end{document}